\documentclass[twocolumn,rnote]{aa} 
\usepackage{graphicx}
\usepackage{natbib}
\usepackage{txfonts}
\def\lax{{$\mathrel{\hbox{\rlap{\hbox{\lower4pt\hbox{$\sim$}}}\hbox{$<$}}}$}}
\def\gax{{$\mathrel{\hbox{\rlap{\hbox{\lower4pt\hbox{$\sim$}}}\hbox{$>$}}}$}}

\begin{document}

   \title{Stellar and Dust Properties of Local Elliptical Galaxies:
Clues to the Onset of Nuclear Activity}
\authorrunning{Zhang et al. }
\titlerunning{Stellar and Dust Properties of Local Elliptical Galaxies}
   \author{Y. Zhang, Q.-S. Gu, and Luis C. Ho}

   \offprints{}

   \institute{
   Department of Astronomy, Nanjing University, Nanjing 210093, China \\
   zhangyu19842007@hotmail.com; qsgu@nju.edu.cn
   \and
   The Observatories of the Carnegie Institution of Washington, 813 Santa Barbara Street, Pasadena, CA 91101, USA   \\
   lho@ociw.edu
   }

   \date{}

  \abstract
    {}
     {We study the stellar and dust properties of a well-defined sample of
     local elliptical galaxies to investigate the relationship between host
     galaxy properties and nuclear activity.}
    {We select a complete sample of 45 ellipticals from the Palomar
     spectroscopic survey of nearby galaxies, which includes 20 low-luminosity
     active galactic nuclei classified as LINERs and 25 inactive galaxies.
     Using a stellar population synthesis method, we
     compare the derived stellar population properties of the LINER versus the
     inactive subsamples.  We also study the dust and stellar surface
     brightness distributions of the central regions of these galaxies using
     high-resolution images obtained with the {\it Hubble Space Telescope}.}
    {Relative to the inactive subsample, ellipticals hosting LINERs share
    similar total optical and near-infrared luminosity, central stellar
    velocity dispersions, and nuclear stellar populations as judged from their
    luminosity-weighted ages and metallicities.  LINERs, on the other hand,
    have a larger fraction of core-type central surface brightness profiles
    and a much higher frequency of circumnuclear dust structures.}
    {Our results support the suggestion that LINERs are powered by
    low-luminosity AGNs rather than by young or intermediate-age stars. Nuclear
    activity in nearby elliptical galaxies seems to occur preferentially in
    those systems where sufficient cold interstellar material has managed to
    accumulate, perhaps via cooling condensations from hot gas.}


   \maketitle
%

\section{Introduction}
 Supermassive black holes (SMBHs) are ubiquitous in elliptical galaxies.
 The discovery of a tight correlation between black hole
 mass and the stellar velocity dispersion of the bulge of the host galaxy
 (\citealt{Gebhardt}; \citealt{Ferrarese}) suggests that the formation and
 growth of central black holes are an integral part of the formation of galactic
 bulges.  However,  the majority of SMBHs in nearby ellipticals are not very active.
In the Palomar spectroscopic survey of nearby galaxies (Ho et al. 1995, 1997a),
 only about 50\% of the ellipticals show detectable emission-line
 nuclei, most of which ($\sim$87\%) are classified as
 low-ionization nuclear emission-line regions (LINERs; \citealt{Heckman}; see Ho 2008
for a review).

The physical origin of LINERs has been intensively debated.
A variety of recent observations, however, provided convincing evidence that
LINERs are low-luminosity active galactic nuclei (AGNs).  This includes the detection
of broad H$\alpha$
emission (Ho et al. 1997b), some of which show double-peaked profiles
(Ho et al. 2000; Shields et al. 2000; Barth et al. 2001) or are
polarized (\citealt{Barth1, Barth2}), ultraviolet variability (Maoz et al.
2005), compact X-ray cores (e.g., Ho et al. 2001; Terashima \& Wilson 2003), and
compact radio cores (e.g., Nagar et al. 2005).  Ho (2008) suggests that nearly
{\it all}\ LINERs are genuine AGNs with low accretion rates.

 The low accretion rates of the SMBHs in nearby ellipticals probably stem from
 the lack of enough cold gas to fuel the
 central engine, as it is generally accepted that nearby elliptical
 galaxies contain mostly old stars and not much cold
 interstellar material. However, apart from the high detection rate of nuclear
optical emission lines in spectroscopic surveys (Phillips et al. 1986;
Ho et al. 1997a), H$\alpha$ + [N~II] narrow-band imaging
 has shown that elliptical galaxies certainly have plenty of extended
 warm ($\sim 10^4$ K) gas (e.g., \citealt{Shields1991}). By coadding images from
 the {\it Infrared Astronomy Satellite}\ survey scans,
\citet{Knapp} show that $\sim$50\% of ellipticals contain
 cool dust. \citet{Temi} find that in 16
 elliptical galaxies observed with the {\it Infrared Space
 Observatory}\ the dust masses are on average 10 times larger
 than those previously estimated from {\it Infrared Astronomy Satellite}\ observations.
Most recently, \citet{Kaneda} even detected polycyclic
 aromatic hydrocarbon emission features in four elliptical
 galaxies with the {\it Spitzer} Infrared Spectrograph. All these
 results suggest that the presence of a significant amount of interstellar material
 in elliptical galaxies is quite common, although its origin is still uncertain.

 We attempt to determine the factor(s) that control the onset of nuclear activity
 by comparing the stellar population, central light distribution, and dust content
 for ellipticals with different levels of nuclear activity drawn from
 the Palomar spectroscopic survey of nearby galaxies (Ho et al. 1995, 1997a).
 This paper is organized as follows.  The sample and analysis method
 are described in Section 2; the results
 of the stellar population synthesis are presented in Section 3;
the discussion and conclusions are given in Section 4.
\section{Sample and Method of Analysis}

The sample for this study was selected from the Palomar optical
spectroscopic survey (Ho et al. 1995, 1997a), which comprises all
nearby galaxies brighter than $B_T=12.5$ mag in the northern
hemisphere.  It presents a fair representation of the local galaxy
population. This is an ideal sample because it is statistically
complete and contains both active ellipticals and inactive
objects, drawn self-consistently from the same parent sample, that
can serve as a control sample. The optical spectra are described
in detail in \citet{Ho2}. Two spectra are available for each
galaxy: the blue section covers $\sim 4230-5110$ \AA\ with a FWHM
spectral resolution of 4 \AA, and the red section covers $\sim
6210-6860$ \AA\ with a FWHM resolution of 2.5 \AA.  We select all
ellipticals from the parent sample classified as either LINERs
{\bf according to the criteria laid out in Ho, Filippenko \&
Sargent (1997a)} or absorption-line (inactive) sources. We reject
NGC~6702 because it only has the red spectrum, and two spheroidals
(NGC 147 and NGC 205) because they have a different formation
history than normal ellipticals (Kormendy 1985). We include NGC
221 (M32), even though its absolute $B$-band magnitude is less
than $-$18 mag, because it shares properties similar to those of
normal ellipticals (\citealt{Kormendy}; \citealt{Ferguson}). Since
the vast majority of the active sources are LINERs, we further
removed 4 Seyferts and 5 transition objects. The final sample,
whose global properties are summarized in Table 1, comprises 20
LINERs and 25 inactive ellipticals.

We modeled the stellar population of our sample using a modified
version of the spectral population synthesis code, {\it
starlight}\footnote{http://www.starlight.ufsc.br/},
(\citealt{Cid04}; \citealt{Gu}). The code does a search for the
best-fitting linear combination of 45 simple stellar populations
(SSPs)---15 ages and 3 metallicities ($0.2\,Z_\odot$,
$1\,Z_\odot$, $2.5\,Z_\odot$)---provided by \citet{BC03} to match
a given observed spectrum $O_\lambda$.  The model SSP spectra
cover 3200 \AA\ to 9500 \AA\ with a FWHM spectral resolution of
$\sim$3 \AA\ (Bruzual \& Charlot 2003, BC03 hereafter). The synthesized spectrum
is given by (\citealt{Cid04}):
\begin{equation}\label{eq1}
    M_\lambda=M_{\lambda_0}\left(\sum^{N_*}_{i=1}x_ib_{i,\lambda}
    r_\lambda\right)\otimes G(v_*,\sigma_*),
\end{equation}

\noindent where $M_\lambda$ is the synthesized spectrum, $M_{\lambda_0}$
is the synthesized flux at normalization wavelength $\lambda_0$ (4750 \AA\ in
this work), $x_i$ is the contribution of the $i$th SSP, $b_{i,\lambda}$ is
the spectrum of the $i$th SSP normalized at $\lambda_0$, and $r_\lambda =
10^{-0.4(A_\lambda - A_{\lambda_0})}$ is the reddening term. The
\citet{Cardelli} extinction law with $R_V = 3.1$ is adopted.  $G(v_*,\sigma_*)$
is the line-of-sight stellar velocity distribution, modeled as a Gaussian
distribution centered at $v_*$ and broadened by dispersion $\sigma_*$.  The
match between the model and the observed spectrum is calculated by
$\chi^2(x, M_{\lambda_0}, A_V, v_*,\sigma_*) = \sum^{N_*}_{i=1}
[(O_\lambda-M_\lambda)w_\lambda]^2$, where the weight spectrum, $w_\lambda$, is
defined as the inverse of the noise of $O_\lambda$. For more details, see
Cid Fernandes et al. (2004, 2005).

In order to account for the possibility of a weak AGN continuum (see Sec. 5.1
of Ho 2008), we add an additional power-law (PL) component ($f_{\nu} \propto
\nu^{-1.5}$) during the fitting. Our fitting results show that, with the
exception of NGC  315 and NGC 3193, the contribution from this PL
component is negligible (less than 2\%).  Strong emission lines, such as
H$\beta$, [O~III] $\lambda\lambda$4959, 5007, H$\alpha$, and
[N~II] $\lambda\lambda$6548, 6583, are masked during the fitting process.
The fitting procedure automatically determines a best-fitting
stellar velocity dispersion ($\sigma_*$), taking into account the instrumental
resolutions of both the Palomar and model library spectra.  Cid Fernandes et
al. (2004) estimate that the $\sigma_*$ values derived from the
code has a typical uncertainty of $\sim$20 km s$^{-1}$.

Figures 1 and 2 show two fitting examples, for the LINER NGC 2768
and the inactive elliptical NGC 4648, respectively. {\bf We have checked
every residual spectrum, and measured the emission line fluxes for
LINERs, as we used the different method to subtract the stellar contribution from
the Palomar survey. We find that the difference of emission line fluxes is less than
10 \%,  which dose not change the spectral classification for our sample.

We also present the stellar extinction (A$_V$) estimated by {\it starlight}  
for each galaxy in Table 1. We find that 6 LINERs and 4 inactive ellipticals 
have negative A$_V$, a problem which is ultimately due  to calibration 
problems in STELIB and mismatches between the metallicity of the 
evolutionary tracks and stellar spectra in BC03. }

\section{Results}

\subsection{Distance, Luminosity, and Stellar Velocity Dispersion}

The distance and absolute blue magnitude of each galaxy, taken
from \citet{HFS}, are listed in Table 1. In Figure 3(a) we show
the cumulative distribution of distance for the two subsamples. We
find that LINERs and inactive ellipticals have similar distance
distributions, with average distances of 29.9 and 25.7 Mpc,
respectively. The LINER subsample contains the most distant object
in the study, NGC  2832 at $D$ = 91.6 Mpc; if we exclude this
object, the average distance for the LINERs becomes 27.6 Mpc. We
use the two-sample Kolmogorov-Smirnov (KS) statistical task {\it
kolmov} in IRAF\footnote{ IRAF is distributed by the National
Optical Astronomy Observatories, which are operated by the
Association of Universities for Research in Astronomy, Inc., under
cooperative agreement  with the National Science Foundation.} to
check whether the two subsamples are drawn from the same parent
distribution. The probability of rejecting the null hypothesis
that the two distributions are the same is $P_{\rm null}$ = 94\%.
Thus, distance effects between the two subsamples will not
introduce any significant biases into our results.

The two subsamples also have very similar total luminosities, both
in the optical and near-infrared bands.  Figure 3(b) shows the
cumulative distributions of absolute blue magnitude.  LINERs have
$M_B$ ranging from $-18.96$ to $-22.24$ mag, with an average value
of $-20.69$ mag, while inactive Es range from $M_B$ = $-15.51$ to
$-21.64$ mag, with an average value of $-20.11$ mag. A KS test
yields $P_{\rm null}$ = 38\%.  We also derive absolute magnitudes
in the $K_s$ band, $\rm M_{K_s}$, using data taken from the Two
Micron All Sky Survey (Skrutskie et al. 2006).  Figure 3(c) shows
that the absolute $K_s$-band magnitudes of LINERs are similar to
those of inactive Es: LINERs have an average $M_{K_s} = -24.54$
mag, to be compared with $M_{K_s} = -23.90$ mag for inactive Es.
The near-infrared luminosity of the active sources have a slight
tendency to be larger than those of the inactive sources, but a KS
test yields $P_{\rm null} \approx 5\%$, which is formally only
barely statistically significant.  Moreover, this conclusion is
not very robust. If we exclude the two most luminous objects from
the active sample (NGC 315 and NGC 2832) and the least luminous
object from the inactive sample (NGC 221), the significance drops
to $P_{\rm null} \approx 14\%$.

Not surprisingly, the similarity between the two subsamples
extends to the stellar velocity dispersions derived from our
synthesis fitting (data in Column 6 of Table 1), whose cumulative
distributions are shown in Figure 4(a). For LINERs, the velocity
dispersion ranges from 144 to 331 km~s$^{-1}$, with an average
value of $\langle \sigma_* \rangle = 251$  km~s$^{-1}$; the
corresponding values for inactive Es are $90 < \sigma_* < 349$
km~s$^{-1}$, with $\langle \sigma_* \rangle = 211$  km~s$^{-1}$. A
KS test for the full sample yields $P_{\rm null} \approx 3\%$, but
the significance drops to $P_{\rm null} \approx 9\%$ after
excluding NGC 315 and NGC 2832 from the LINER sample and NGC 221
from the inactive sample.

\subsection{Stellar Populations}

As mentioned in Section 2, our spectral library covers 15 ages and 3
metallicities, which produce a total of 45 SSPs. Here we rebin the 45 SSPs
into 5 components according to age: I ($10^6 \leqslant t < 10^7$ yr), II
($10^7 \leqslant t < 10^8$ yr), III ($10^8 \leqslant t < 10^9$ yr),
IV ($10^9 \leqslant t < 10^{10}$ yr), and V ($t \geqslant 10^{10}$ yr).
We derive the light-weighted mean stellar age using the formula in
\citet{Cid05}:

\begin{equation}\label{eq2}
    \langle \log t_*\rangle_L=\sum^{45}_{i=1}x_i\log t_i,
\end{equation}

\noindent where $x_i$ denotes the contribution of the $i$th SSP and $t_i$ its
corresponding age.  Similarly, the
light-weighted mean stellar metallicity is computed from

\begin{equation}\label{eq3}
    \langle Z_*\rangle_L=\sum^{45}_{i=1}x_iZ_i,
\end{equation}

\noindent where $x_i$ represents the contribution of the $i$th SSP with
metallicity $Z_i$.
The synthesis results are presented in Table 2. Columns (2) through (6) give
the fractional light contribution from the AGN PL component and from stellar
populations I--V.  Columns (7) and (8) list the mean stellar age and
metallicity.

Figure 5 shows the contribution of each stellar population for the
LINER (upper panel) and inactive E (lower panel) subsamples. It is
clear that young and intermediate-age stellar populations
contribute very little to the total flux for either group.  For
LINERs, stellar population I, II, and III contribute only $<$1\%,
2\%, and $<$1\%, respectively; for inactive Es, the corresponding
values are 1\%, 2\%, and $<$1\%. Most of the optical light is
dominated by an old population. The mean contribution from stellar
population IV and V are 11\% and 86\% for LINERs, and 20\% and
76\% for inactive Es.

Figure 4(b) shows the cumulative distribution of mean stellar age
for LINERs and inactive Es.  Their average age is 8.1 Gyr and 7.1
Gyr, respectively. The only exception is the LINER NGC 4374, which
has a dominant 1.4 Gyr old stellar population that contributes
89\% of the light. Our stellar population synthesis suggests that
LINERs and inactive Es have very similar stellar populations, both
lacking young and intermediate-age populations. A KS test yields
$P_{\rm null} = 31\%$(68\%, NGC221 excluded).  The similarity of
the stellar populations of the two groups further extends to the
derived metallicities (Fig. 4(c)): the average metallicity for
both subsamples is $\sim 1.5\,Z_\odot$.

\subsection{Central Light Profile}

Central surface brightness profiles carry many clues on the formation and
evolutionary history of the galaxy (Faber et al. 1997; \citealt{Lauer05}). We
compare the surface brightness profiles of LINERs and inactive Es derived from
{\it Hubble Space Telescope (HST)}\ images, and try to find whether there is
any difference between the two subsamples.  The high angular resolution of
{\it HST}\ is crucial to derive robust central profiles of galaxies.

\citet{Lauer95} classified the central surface brightness profiles of early-type
galaxies into two types.  Core galaxies have central light distributions
that turnover from a steep outer profile to a shallower cusp interior to some
break radius.  Power-law galaxies, on the other hand, display continuously
rising inner profiles down to the resolution of {\it HST}.  The two classes
correlate with the large-scale properties of the galaxies.  Core galaxies are
invariably luminous giant ellipticals that are slowly rotating and show boxy
isophotes; power-law galaxies occupy more intermediate luminosities, rapidly
rotate, and show disky isophotes.  The physical implications of these two
classes for their formation histories are discussed in detail in Faber et al.
(1997).

Published {\it HST}\ surface brightness profiles and
classifications are available for most of our sample, as
documented in Table 1. Among the 16 LINERs with profile
classifications, 14 (88\%) are core galaxies; only one galaxy, NGC
4494, is classified as a power-law system, and another, NGC 7626,
has an intermediate profile.  Four (NGC 315, 2768, 3226, and 5322)
are too dusty to determine their profiles reliably and cannot be
classified. However, for the inactive galaxies, only 12 out of 22
(55\%) have cores, whereas 9 (41\%) have power-law profiles; NGC
821 has an intermediate classification, and three others lack
data.

\subsection{Central Dust Content}

Elliptical galaxies are traditionally viewed as structurally simple systems
composed of old stars and lacking in cold gas and dust.
Recent observations, especially from {\it HST}\ images,  show
unambiguous evidence that many ellipticals exhibit central dust structures
(\citealt{Dokkum}; \citealt{Tran}; \citealt{Lauer05}).  \citet{Lauer05}
suggested a dust-settling sequence in early-type galaxies: the dust starts off
poorly organized, then gradually settles down to the center and forms dusty
rings or disks.  The origin of gas and dust in elliptical galaxies is not well
understood.  Two possibilities are frequently discussed---an internal origin
from stellar mass loss or an external origin from accretion of a gas-rich
companion (\citealt{Lauer05}; \citealt{Sarzi}). Although we do not have
detailed kinematic information on our galaxies, they are morphologically
regular and do not show obvious signs of recent interactions or mergers.

We compare the dust properties of LINERs and inactive Es in our
sample. Information on the central dust properties, gathered from
a literature search, is summarized in Table 1. Among the LINERs,
16 out of 20 (80\%) have detected central dust structures.  In
stark contrast, only 2 out of 22 (9\%) {\bf inactive ellipticals}
with suitable {\it HST}\ data show dust features.

{\bf Since LINER hosts seems more dusty than inactive Es, we 
would expect LINERs to have more stellar extinction than inactive Es.  We thus
correlated the stellar extinction (A$_V$) with the YEs/No dust-content
information. After removing the negative A$_V$, we find there is no significant 
difference in A$_V$, the average of 
A$_V$ is 0.26 mag and 0.22 mag, for LINERs and inactive ellipticals, respectively. }

\section{Discussion and Conclusions}
Analyzing a statistically complete sample of 45 elliptical galaxies from the
Palomar spectroscopic survey of nearby galaxies, comprising 20 LINERs and
25 inactive members, we show that galaxies hosting LINER nuclei, compared to
systems lacking optical evidence for nuclear activity, have very similar
stellar properties, both on global and nuclear scales.  Specifically, we find
that the two galaxy subgroups have essentially indistinguishable total optical
and near-infrared luminosities, central stellar velocity dispersions, and
luminosity-weighted mean stellar ages and metallicities on nuclear (\lax
300 pc) scales.  The lack of young or intermediate-age stars in LINERs strongly
rule out starburst and post-starburst models for the excitation of LINER-like
emission, further strengthening the proposition that LINERs are low-luminosity
AGNs (Ho 2008).

Our results are qualitatively consistent with those from previous
studies of the Palomar survey.  \citet{Ho6} systematically
investigated the stellar population of the emission-line nuclei in
the Palomar survey and found that LINERs generally contain evolved
stellar populations.  Other evidence concerning the old ages of
the nuclear stellar population in nearby low-luminosity AGNs is
summarized in Ho (2008).  Ho et al., however, did not derive
detailed ages or metallicities from a population synthesis
analysis.  This work is the first quantitative analysis of its
kind for the Palomar survey.  On the other hand, studies of larger
samples of LINERs drawn from the Sloan Digital Sky Survey reach
somewhat different conclusions.  While LINERs on average have
older stars than Seyferts (Kewley et al. 2006), Graves et al.
(2007) and Schawinski et al.  (2007) show that LINERs have a
tendency to exhibit somewhat younger populations compared to
inactive early-type galaxies devoid of emission lines. {\bf The
difference may arise from the following two points. First, the sources from the
Sloan Digital Sky Survey have redshifts $z \approx
0.05-0.1$, much more distant than the Palomar galaxies. As
discussed by Ho (2008), aperture effects render the interpretation
of LINER emission in distant galaxies particularly ambiguous.  Second,
the stellar population analysis method of this paper is different from
that used in SDSS galaxies.
However, the reported age differences are
relatively small and may require more data than we have and/or a more 
detailed analysis to be checked.}

Surprisingly, {\bf we find that} LINER host galaxies have a
greater tendency to exhibit central core light profiles.  Lauer et
al. (2005) suggest that core galaxies are more luminous, more
massive, and may be older than power-law galaxies, seemingly at
odds with our results.  Considering the modest size of our sample,
however, and the large scatter in the statistical correlations
between profile type and global properties, this disagreement is
not too alarming.

By far the clearest distinction between active and inactive ellipticals is
that the former has a much higher likelihood of containing circumnuclear dust.
This finding is in excellent agreement with that of \citet{Lauer05}, who also
found a strong correlation between the incidence of nebular line
emission---which in ellipticals invariably signifies a LINER classification
(Ho 2008)---and central dust features: 90\% of emission-line nuclei show
detected central dust structures, to be compared with just 4\% for the
lineless systems.  \citet{Ravindranath} similarly suggested a connection
between H$\alpha$ emission and dust features, and the tendency for LINERs to
be more dusty.  Recently, Sim\~{o}es Lopes et al. (2007) presented  strong
evidence that circumnuclear dust is correlated with AGN activity in early-type
galaxies.  Among 34 early-type galaxies classified as AGNs, all show
circumnuclear dust structures, to be compared with only 26\%  in the control
sample of inactive systems.

Taken collectively, the above evidence suggests that ellipticals with weakly
active, LINER nuclei are more gas-rich than their inactive counterparts. Given
the lower stellar density in the cores, we actually expect {\it less}\ cold gas
accumulation from stellar mass loss, not more, among the LINER hosts (see,
e.g., Soria et al. 2006).  On the other hand, massive ellipticals are
surrounded by hot X-ray halos (e.g., Jones et al. 2002), whose cooling time is
sufficiently short that cooling flows should develop.  We speculate that the
central dust features in core-type LINER ellipticals may partly originate from
cooling condensations in the X-ray-emitting gas.  The ``extra'' cold gas and
dust, in combination with direct spherical accretion from the hot gas
itself, may be responsible for sustaining the weak nuclear activity.
Perhaps elliptical galaxies with LINER nuclei identify the subclass of
early-type galaxies in which this mode of accretion is significant.

\begin{acknowledgements}
We thank an anonymous referee for comments that led
to significant improvements in the paper, and Roberto Cid Fernandes for
providing us the updated code, {\it starlight}, for stellar population synthesis.
This work is supported by Program for New Century Excellent Talents
in University (NCET), the National Natural Science Foundation of China
under grants 10221001 and 10633040, and  the National Basic Research
Program (973 program No. 2007CB815405).  This research has made use of NASA's
Astrophysics Data System Bibliographic Services and the NASA/IPAC Extragalactic Database (NED), which is operated by the Jet Propulsion Laboratory,
California Institute of Technology, under contract with the National Aeronautics
and Space Administration. This publication makes use of data products
from the Two Micron All Sky Survey, which is a joint project of the University
of Massachusetts and the Infrared Processing and Analysis Center/California
Institute of Technology, funded by the National Aeronautics and Space
Administration and the National Science Foundation.
\end{acknowledgements}

\clearpage

\begin{table}
\caption{Global and central properties}             
\begin{tabular}{c c c c c c c c c c c}        
\hline\hline                 
LINERs & $D$ (Mpc) & $M_B$ (mag) & $M_{K_s}$ (mag) & Class & $\sigma_*$ (km s$^{-1}$) & A\_V (mag)& Profile & Ref. & Dust Properties & Ref. \\
\hline                        
NGC 315  & 65.8 & $-$22.22 & $-$26.14 & L1.9 & 322 &   0.42&       &      &      Dust disk      & 2\\
NGC 1052 & 17.8 & $-$19.90 & $-$23.80 & L1.9 & 240 &   0.04&  core & 1,9  &      Has dust       & 3  \\
NGC 2768 & 23.7 & $-$21.17 & $-$24.88 & L2   & 170 &   0.32&       &      & Spiral dust lanes   & 4\\
NGC 2832 & 91.6 & $-$22.24 & $-$26.11 & L2:: & 329 &   0.02&  core & 1,10 &       No dust       & 5\\
NGC 3193 & 23.2 & $-$20.10 & $-$23.85 & L2:  & 193 &$-$0.02&  core &  1,8 &       No dust       & 6\\
NGC 3226 & 23.4 & $-$19.40 & $-$23.28 & L1.9 & 227 &   0.61&       &      &      Dust disk      & 2\\
NGC 3379 & 8.1  & $-$19.36 & $-$23.27 & L2/T2:: & 237 &$-$0.17&  core &  1,4 &  Dust ring or disk  & 4\\
NGC 3608 & 23.4 & $-$20.16 & $-$23.75 & L2/S2: & 218 &$-$0.10&  core &  1,4 &  Dust ring or disk  & 4\\
NGC 4261 & 35.1 & $-$21.37 & $-$25.46 & L2   & 331 &   0.16&  core &  1,3 &      Dust disk      & 2\\
NGC 4278 & 9.7  & $-$18.96 & $-$22.75 & L1.9 & 269 &$-$0.11&  core &  1,4 & Spiral dust lanes   & 4\\
NGC 4374 & 16.8 & $-$21.12 & $-$24.90 & L2  & 246 &   0.96&  core &  1,3 &      Dust lanes     & 3\\
NGC 4486 & 16.8 & $-$21.64 & $-$25.31 & L2   & 328 &   0.04&  core & 1,11 &       No dust       & 7\\
NGC 4494 & 9.7  & $-$19.38 & $-$22.94 & L2:: & 144 &   0.14&   PL  &  1,8 &  Dust ring or disk  & 4\\
NGC 4589 & 30.0 & $-$20.71 & $-$24.63 & L2   & 243 &   0.34&  core &  1,4 &Chaotic dust patches & 4\\
NGC 4636 & 17.0 & $-$20.72 & $-$24.73 & L1.9 & 215 &   0.14&  core &  1,3 &      Dust lanes     & 7\\
NGC 5077 & 40.6 & $-$20.83 & $-$24.83 & L1.9 & 266 &   0.04&  core &  1,8 &     Dust filaments  & 6,8\\
NGC 5322 & 31.6 & $-$21.46 & $-$25.34 & L2:: & 232 &$-$0.08&       &      &   Dust ring or disk & 4\\
NGC 5813 & 28.5 & $-$20.85 & $-$24.86 & L2:  & 257 &   0.21&  core &  1,4 &       Dust disk     & 6\\
NGC 5982 & 38.7 & $-$20.89 & $-$24.79 & L2:: & 267 &$-$0.12&  core &  1,4 &        No dust      & 2,3,6\\
NGC 7626 & 45.6 & $-$21.23 & $-$25.26 & L2:: & 286 &   0.22&  int  &  1,3 &   Dust ring or disk & 2,4\\
average & 29.9 & $-$20.69  & $-$24.54 &      & 251 &   0.15&       &      &                     &   \\
rms     & 20.1 &     0.92   &  0.99   &      &  50 &   0.28&       &      &                     &   \\
\hline                                   
Inactive Es& $D$ (Mpc) & $M_B$ (mag) & $M_{K_s}$ (mag) & Class & $\sigma_*$ (km s$^{-1}$) & A\_V (mag)& Profile & Ref. & Dust Properties & Ref. \\
\hline                        
NGC 221  &  0.7 & $-$15.51 & $-$19.13 & A &  90 &   0.48&  PL   & 3,12  &    No dust      &  3,12\\
NGC 821  & 23.2 & $-$20.11 & $-$23.93 & A & 205 &   0.31&  int  &  1,4  &    No dust      &  3,4\\
NGC 2634 & 30.2 & $-$19.69 & $-$23.14 & A & 193 &   0.18&  PL   &  1,8  &    No dust      &  6\\
NGC 3348 & 37.8 & $-$21.05 & $-$24.92 & A & 221 &   0.32&  core &  1,8  &    No dust      &  6\\
NGC 3377 & 8.1  & $-$18.47 & $-$22.10 & A & 166 &   0.36&  PL   &  1,4  & Dust filaments  &  4,6\\
NGC 3610 & 29.2 & $-$20.79 & $-$24.42 & A & 172 &   0.06&  PL   &  1,4  &    No dust      &  4\\
NGC 3613 & 32.9 & $-$20.93 & $-$24.58 & A & 219 &   0.19&  core &  1,8  &    No dust      &  4\\
NGC 3640 & 24.2 & $-$20.73 & $-$24.40 & A & 196 &   0.27&  core &  1,4  &    No dust      &  4,6\\
NGC 4291 & 29.4 & $-$20.09 & $-$23.92 & A & 300 &   0.04&  core &  1,4  &    No dust      &  3,4,6\\
NGC 4339 & 25.5 & $-$19.72 & $-$23.50 & A & 138 &$-$0.01&       &       &                 &    \\
NGC 4365 & 16.8 & $-$20.64 & $-$24.49 & A & 251 &   0.30&  core &  1,4  &    No dust      &  4,6\\
NGC 4406 & 16.8 & $-$21.39 & $-$25.02 & A & 233 &   0.22&  core &  1,4  &    No dust      &  4\\
NGC 4473 & 16.8 & $-$20.10 & $-$23.97 & A & 176 &   0.18&  core &  1,4  &    No dust      &  4\\
NGC 4478 & 16.8 & $-$18.92 & $-$22.77 & A & 142 &   0.21&  core &  1,4  &    No dust      &  4,6\\
NGC 4564 & 16.8 & $-$19.17 & $-$23.19 & A & 160 &   0.24&  PL   &  1,8  &    No dust      &  6,8\\
NGC 4621 & 16.8 & $-$20.60 & $-$24.38 & A & 245 &   0.14&  PL   &  1,4  &    No dust      &  4,6\\
NGC 4648 & 27.5 & $-$19.49 & $-$23.35 & A & 222 &   0.26&  PL   &  1,8  &   Dust disk     &  6,8\\
NGC 4649 & 16.8 & $-$21.43 & $-$25.39 & A & 349 &   0.23&  core &  1,4  &    No dust      &  4\\
NGC 4660 & 16.8 & $-$19.06 & $-$22.92 & A & 198 &   0.27&  PL   &  1,4  &    No dust      &  4\\
NGC 4914 & 62.4 & $-$21.64 & $-$25.33 & A & 244 &$-$0.04&       &       &                 &    \\
NGC 5557 & 42.6 & $-$21.17 & $-$25.07 & A & 290 &$-$0.20&  core &  1,4  &    No dust      &  4,6\\
NGC 5576 & 26.4 & $-$20.34 & $-$24.28 & A & 211 &   0.08&  core &  1,4  &    No dust      &  4,6\\
NGC 5638 & 28.4 & $-$20.21 & $-$24.01 & A & 152 &$-$0.32&       &       &                 &    \\
NGC 5831 & 28.5 & $-$19.96 & $-$23.84 & A & 167 &   0.13&  PL   &  1,8  &    No dust      &  6,8\\
NGC 7619 & 50.7 & $-$21.60 & $-$25.49 & A & 337 &   0.24&  core &  1,4  &    No dust      &  4\\
average  & 25.7 & $-$20.11 & $-$23.90 &   & 211 &   0.17&       &       &                 &    \\
rms      & 13.1 &    1.29  &   1.32   &   &  62 &   0.17
&       &       &                 &    \\
\hline                                   

\end{tabular}
\begin{list}{}{}
\item[] Col. (5): Nuclear spectral classification from the Palomar survey
(Ho et al. 1997a): L = LINER, S = Seyfert, T = transition object, and A =
absorption-line nucleus (inactive).
The last two rows show the average value and rms of each column.
Ref: 1. \citealt{Lauer07}; 2. \citealt{Gonzalez}; 3.
\citealt{Ravindranath}; 4. \citealt{Lauer05}; 5. \citealt{Martel};
6. \citealt{Tran}; 7. \citealt{Dokkum}; 8. \citealt{Rest}; 9.
\citealt{Quillen}; 10. \citealt{Laine}; 11. \citealt{Lauer95}; 12.
\citealt{Faber}.
\end{list}
\end{table}

\clearpage

\begin{table}
\caption{Synthesis results}             
\label{table:2}      
\centering                          
\begin{tabular}{c c c c c c c c c}        
\hline\hline                 
LINERs & PL (\%) & I (\%) & II (\%) & III (\%) & IV (\%) & V (\%) & log Age (yr) & Metallicity $(Z_\odot)$ \\
\hline                        
NGC 315    &    7        &    2        & $<$1        & $<$1        & $<$1        &   92      &   9.79  &  1.29\\
NGC 1052   & $<$1        & $<$1        & $<$1        & $<$1        &   11        &   89      &  10.03  &  1.53\\
NGC 2768   & $<$1        & $<$1        & $<$1        & $<$1        &   37        &   63      &   9.79  &  2.07\\
NGC 2832   & $<$1        & $<$1        & $<$1        & $<$1        & $<$1        &   99      &  10.11  &  1.29\\
NGC 3193   &    9        & $<$1        & $<$1        & $<$1        & $<$1        &   91      &   9.72  &  1.12\\
NGC 3226   & $<$1        & $<$1        & $<$1        & $<$1        &   72        &   28      &   9.79  &  1.00\\
NGC 3379   & $<$1        & $<$1        & $<$1        & $<$1        & $<$1        &   99      &  10.11  &  1.68\\
NGC 3608   &    2        & $<$1        & $<$1        & $<$1        & $<$1        &   98      &  10.05  &  1.23\\
NGC 4261   & $<$1        & $<$1        & $<$1        & $<$1        &    5        &   95      &  10.06  &  1.49\\
NGC 4278   & $<$1        &    1        & $<$1        & $<$1        & $<$1        &   99      &  10.07  &  1.75\\
NGC 4374   & $<$1        & $<$1        &    2        & $<$1        &   90        &    8      &   9.18  &  2.46\\
NGC 4486   & $<$1        &    2        &    1        & $<$1        & $<$1        &   96      &   9.99  &  0.86\\
NGC 4494   & $<$1        & $<$1        &    7        & $<$1        &   11        &   83      &   9.80  &  1.25\\
NGC 4589   & $<$1        & $<$1        &   11        & $<$1        &    1        &   88      &   9.76  &  2.50\\
NGC 4636   & $<$1        & $<$1        & $<$1        & $<$1        & $<$1        &   99      &  10.04  &  1.00\\
NGC 5077   & $<$1        &    4        & $<$1        & $<$1        & $<$1        &   96      &   9.96  &  1.74\\
NGC 5322   & $<$1        & $<$1        &   12        & $<$1        & $<$1        &   88      &   9.75  &  2.50\\
NGC 5813   & $<$1        & $<$1        & $<$1        & $<$1        & $<$1        &   99      &  10.11  &  1.36\\
NGC 5982   &    1        & $<$1        & $<$1        & $<$1        & $<$1        &   98      &   9.98  &  0.99\\
NGC 7626   & $<$1        & $<$1        & $<$1        & $<$1        & $<$1        &   99      &  10.11  &  1.25\\
average    &    1.0      &    0.4      &    1.7      & $<$1        &   11        &   86      &   9.91  &  1.52\\
rms        &    2.5      &    1.0      &    3.6      & $<$1        &   25        &   24      &   0.22  &  0.51\\
\hline                                   
Inactive Es &   & I (\%) & II (\%) & III (\%) & IV (\%) & V (\%) & log Age (yr)& Metallicity $(Z_\odot)$\\
\hline                        
NGC 221    &             &    1        & $<$1        &    4        &   94        & $<$1       & 9.33   &  1.88\\
NGC 821    &             & $<$1        & $<$1        & $<$1        & $<$1        &   99       & 10.10  &  1.18\\
NGC 2634   &             &    1        & $<$1        & $<$1        &    1        &   98       & 10.07  &  1.28\\
NGC 3348   &             & $<$1        &    4        & $<$1        & $<$1        &   96       & 10.01  &  2.50\\
NGC 3377   &             & $<$1        &    8        & $<$1        &   11        &   82       &  9.34  &  2.49\\
NGC 3610   &             & $<$1        &    7        & $<$1        &   56        &   37       &  9.36  &  2.40\\
NGC 3613   &             & $<$1        &    3        & $<$1        &   24        &   73       &  9.80  &  1.57\\
NGC 3640   &             & $<$1        & $<$1        & $<$1        &   30        &   70       &  9.82  &  1.45\\
NGC 4291   &             &    4        &    2        & $<$1        &   29        &   65       &  9.70  &  2.42\\
NGC 4339   &             & $<$1        &    3        & $<$1        &    1        &   95       &  9.96  &  1.59\\
NGC 4365   &             & $<$1        &   12        & $<$1        & $<$1        &   88       &  9.75  &  1.28\\
NGC 4406   &             &    2        & $<$1        & $<$1        & $<$1        &   98       & 10.04  &  0.94\\
NGC 4473   &             & $<$1        & $<$1        & $<$1        &   29        &   71       &  9.85  &  1.49\\
NGC 4478   &             & $<$1        & $<$1        & $<$1        &   99        & $<$1       &  9.66  &  0.90\\
NGC 4564   &             &    3        & $<$1        & $<$1        & $<$1        &   97       & 10.00  &  1.10\\
NGC 4621   &             &    2        & $<$1        & $<$1        & $<$1        &   98       & 10.03  &  1.18\\
NGC 4648   &             & $<$1        & $<$1        & $<$1        & $<$1        &   99       & 10.11  &  1.23\\
NGC 4649   &             & $<$1        & $<$1        & $<$1        &    3        &   96       & 10.03  &  0.95\\
NGC 4660   &             & $<$1        &    5        & $<$1        &   15        &   80       &  9.76  &  1.23\\
NGC 4914   &             & $<$1        &    8        & $<$5        &   87        & $<$1       &  9.45  &  2.43\\
NGC 5557   &             & $<$1        & $<$1        & $<$1        &    1        &   99       & 10.08  &  1.06\\
NGC 5576   &             &    2        &    6        & $<$1        &   25        &   66       &  9.62  &  1.52\\
NGC 5638   &             & $<$1        & $<$1        & $<$1        & $<$1        &   99       & 10.11  &  1.16\\
NGC 5831   &             & $<$1        & $<$1        & $<$1        & $<$1        &   99       & 10.11  &  1.22\\
NGC 7619   &             & $<$1        & $<$1        & $<$1        & $<$1        &   99       & 10.11  &  1.24\\
average    &             &    0.6      &    2.3      &    0.4      &   20        &   76       &  9.85  &  1.51\\
rms        &             &    1.1      &    3.4      &    1.3      &   31        &   32       &  0.26  &  0.53\\
\hline                                   

\end{tabular}
\begin{list}{}{}
\item[] Cols. (2)--(7) list the fractional
contribution to the total flux by the PL and by stellar populations I, II,
III, IV, and V, respectively. Contributions less than 1\% are listed as upper limits.
Col. (8) gives the mean stellar age, and Col. (9) gives the mean metallicity.
The last two rows show the average value and rms of each column.
\end{list}
\end{table}
\clearpage

   \begin{figure}
   \centering
   \includegraphics[width=\columnwidth]{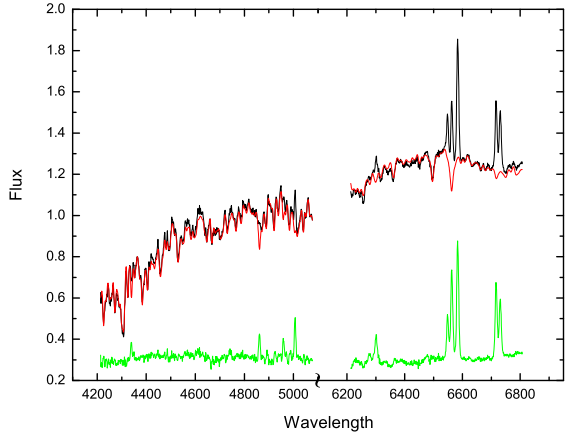}
      \caption{Synthesis results for the LINER NGC 2768.
      The flux is normalized at 4750 \AA. The black line
      is the observed spectrum; the red line is the synthesized
      spectrum; and the green line is the residual spectrum
      after subtracting the synthesized spectrum from the observed spectrum.
      The vertical scale has been shifted arbitrarily.
      There are no data in the region $\sim 5110-6210$ \AA.
              }
         \label{}
   \end{figure}

   \begin{figure}
   \centering
   \includegraphics[width=\columnwidth]{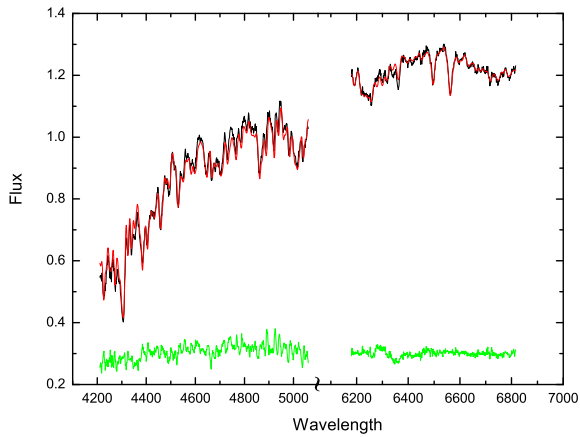}
      \caption{Same as Figure 1, but for the inactive elliptical NGC 4648.
              }
         \label{}
   \end{figure}

   \begin{figure}
   \centering
   \includegraphics[width=\columnwidth]{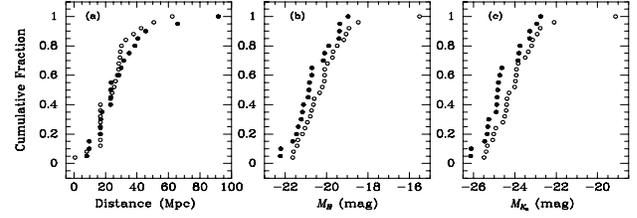}
      \caption{Cumulative distribution of (a) distance, (b) absolute blue magnitude, (c) absolute $K_s$-band magnitude
       for LINERs (filled circles) and inactive ellipticals (open circles).
              }
         \label{}
   \end{figure}

   \begin{figure}
   \centering
   \includegraphics[width=\columnwidth]{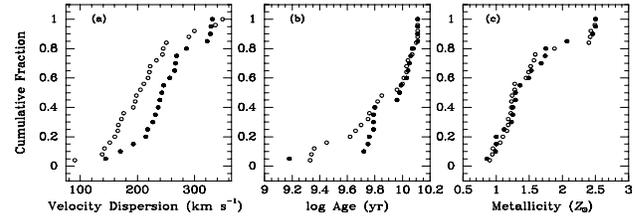}
      \caption{Cumulative distribution of (a) central stellar velocity dispersion, (b) mean stellar age, (c) mean stellar metallicity
       for LINERs (filled circles) and inactive ellipticals (open circles).
              }
         \label{}
   \end{figure}

   \begin{figure}
   \centering
   \includegraphics[width=\columnwidth]{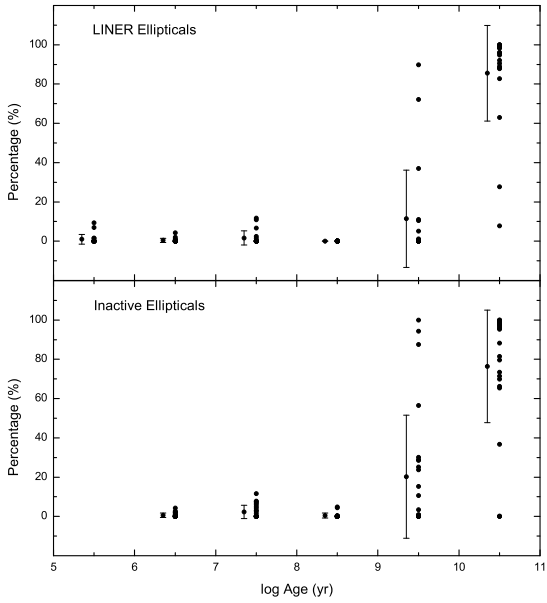}
      \caption{The fractional contribution to the total flux of each stellar
      population for LINERs (upper panel) and inactive ellipticals
      (lower panel). The abscissa gives the logarithm of the age, where 5.5 corresponds to the AGN PL component;
       6.5, 7.5, 8.5, 9.5, and 10.5 correspond to stellar populations I, II, III,
      IV, and V, respectively. The point
and error bar to the left of each population indicate the average fraction and rms, respectively.  We did not fit a PL component to the inactive ellipticals.
              }
         \label{}
   \end{figure}

\end{document}